\documentclass[aip,jap,graphicx,reprint]{revtex4-1}

\usepackage{amssymb}
\usepackage{amsmath}
\usepackage{bm}
\usepackage{graphicx}
\usepackage{epstopdf}

\draft 

\begin{document}

\title{Investigations of afterpulsing and detection efficiency recovery in superconducting nanowire single-photon detectors}

\author{Viacheslav Burenkov}
\email[]{viacheslav.burenkov@utoronto.ca}
\affiliation{Department of Physics, University of Toronto, Toronto, Ontario, M5S 1A7, Canada}

\author{He Xu}
\affiliation{Department of Electrical and Computer Engineering, University of Toronto, Ontario, M5S 3G4, Canada}

\author{Bing Qi}
\affiliation{Department of Electrical and Computer Engineering, University of Toronto, Ontario, M5S 3G4, Canada}
\affiliation{Center for Quantum Information and Quantum Control, University of Toronto, Toronto, Ontario, Canada}

\author{Robert H. Hadfield}
\affiliation{School of Engineering, University of Glasgow, Glasgow, G12 8QQ, United Kingdom}

\author{Hoi-Kwong Lo}
\affiliation{Department of Physics, University of Toronto, Toronto, Ontario, M5S 1A7, Canada}
\affiliation{Department of Electrical and Computer Engineering, University of Toronto, Ontario, M5S 3G4, Canada}
\affiliation{Center for Quantum Information and Quantum Control, University of Toronto, Toronto, Ontario, Canada}


\begin{abstract}
We report on the observation of a non-uniform dark count rate in Superconducting Nanowire Single Photon Detectors (SNSPDs), specifically focusing on an afterpulsing effect present when the SNSPD is operated at a high bias current regime. The afterpulsing exists for real detection events (triggered by input photons) as well as for dark counts (no laser input). In our standard set-up, the afterpulsing is most likely to occur at around $180$~ns following a detection event, for both real counts and dark counts. We characterize the afterpulsing behavior and speculate that it is not due to the SNSPD itself but rather the amplifiers used to boost the electrical output signal from the SNSPD. We show that the afterpulsing indeed disappears when we use a different amplifier with a better low frequency response. We also examine the short-lived enhancement of detection efficiency during the recovery of the SNSPD due to temporary perturbation of the bias and grounding conditions.
\end{abstract}


\maketitle 


\section{Introduction}
\label{Section1}

Superconducting Nanowire Single Photon Detectors (SNSPDs) are a relatively new technology for detecting single infrared photons \cite{Goltsman2001, Natarajan2012}. They can offer certain advantages over other single photon detectors due to their potentially short dead time, small timing jitter \cite{Dauler2009} and low dark count rate \cite{Goltsman2005}. As such, they have been used in many areas of research, including quantum key distribution (QKD) \cite{Bennett1984, Ekert1991, Scarani2009, Hadfield2006, Takesue2007, Stucki2009}, quantum state tomography \cite{Cramer2010} and other quantum optic experiments. There has been growing interest in completely characterizing these quantum detectors through a process called detector tomography \cite{Lundeen2008, Akhlaghi2011, Natarajan2013}. Yet, such characterization is often based on the response of a detector to a one-shot input, when the initial state of a detector is ``active''. In other words, it does not consider the possibility that a detector is already ``dead'' due to an earlier detection event (a ``click'') and the fact that the response of a detector to a signal actually depends on its initial state and when the last detection event occurs.

In the context of QKD, we remark that the existing theoretical models for photon detectors are often too simplistic and do not take into account various imperfections in practical detectors. This is highly undesirable because, for one thing, those imperfections may open up security loopholes which allow Eve to hack commercial QKD systems, as demonstrated by the recent quantum hacking experiments against InGaAs APDs \cite{Zhao2008, Lydersen2010} and an SNSPD \cite{Lydersen2011}. Therefore, it is important to re-examine existing models for detectors and see if they describe practical detectors well.

Regarding SNSPDs, two assumptions are often made. First, it is commonly assumed without proof that the dark counts (spurious clicks that occur with no light input) of an SNSPD are uniform in time. Secondly, initial experimental studies \cite{Kerman2006} indicate that detection efficiency recovers continuously, mirroring the recovery of the bias current in the nanowire \cite{Yang2007}.

In this paper, we find that, rather surprisingly, both assumptions are \textit{invalid} for a practical SNSPD. Our investigation consists of three parts. In the first part of our investigation, with no laser input, we study the distribution of dark counts. We find that, rather unexpectedly, dark counts are not uniform. In fact, dark counts show a clustering effect, which we refer to as ``afterpulsing''. More concretely, for our SNSPD, the total dark count probability could be separated into a uniformly distributed ``pure'' dark count probability and a highly time-dependent ``extra'' dark count probability, occurring on a time scale of around $180$~ns after a previous dark count.

We found that the probability of afterpulsing of dark counts increases exponentially as the bias current is increased. At high bias currents, afterpulsing of dark counts can be an important contribution to the overall dark count rate. Besides, at high bias currents, more than one afterpulse can occur, with a time interval of about $180$~ns between the adjacent afterpulses. At very high bias current, a long train of afterpulses can occur. We study the distribution of the number of pulses as a function of bias current. We discuss the implications of our finding on the security and performance of QKD with SNSPDs.

In the second part of our investigation, with a pulsed laser on with a repetition rate of $2$~MHz, we study the distribution of detection events. We observe that about 180 ns following a real detection event due to an input pulse, there is an enhanced probability for our SNSPD to register a (spurious) detection event. Such a spurious detection event is commonly called an afterpulse.  We note that two papers related to afterpulsing in SNSPDs (with laser on) have been published recently \cite{Fujiwara2011, Marsili2012}. In Ref. \cite{Fujiwara2011}, the afterpulsing effect of an SNSPD with a laser input has been studied on the time scale of $100$~ns. Here, we study the afterpulsing effect with a laser input at a much larger time-scale (of $1000$~ns) and, for the first time, report the secondary afterpulsing effect (the afterpulse of an afterpulse). Indeed, we find that about $180$~ns after the first afterpulse, there is an enhanced probability of having a second afterpulse. This is similar to our finding in the first part of our investigation (with no laser input).

In the third part of our investigation, we study the recovery curve of SNSPD after a detection event. Here, we report an unexpected recovery of detection efficiency after a detector click. Contrary to the widespread belief, the detection efficiency does not monotonically rise to its nominal value but instead increases beyond it on the same time scale as the afterpulsing, before dropping back to the nominal value.

The paper is structured as follows. Sec.~II describes our experimental set-up in detail and defines what exactly constitutes a detection event. Sec.~\ref{Section4} presents our results of a non-uniform distribution of the dark count events in the SNSPD, and introduces the aforementioned afterpulse effect. In Sec.~\ref{Section5}, we characterize the afterpulsing effect and explain it as a reflection in the readout circuit used, as previously proposed in Ref. \cite{Fujiwara2011}. Analysis of afterpulsing is then extended to experiments with incoming light in Sec.~\ref{Section6}. In Sec.~\ref{Section7} we present our results of detection efficiency recovery following a detection event, showing an unexpected temporary rise in detection efficiency beyond the nominal value. In Sec.~\ref{Section8} we demonstrate experimentally that afterpulsing can be eliminated by using a different amplifier. Finally, we make a summary and concluding comments in Sec.~\ref{Section9}, including highlighting the significance of using an SNSPD with these properties for QKD.

\section{Experimental set-up of SNSPD}
\label{Section2}

The SNSPD consists of a superconducting nanowire, which is current-biased just below its critical current. Photon detection is based on the fact that an incoming photon can induce a transition of the nanowire from its superconducting state to a resistive state, resulting in a voltage pulse that can then be amplified and detected \footnote{See supplementary material at [URL will be inserted by AIP] for the outline of the detection process.}.

The SNSPD used in this study is based on a single $100$~nm width nanowire covering a $20$~$\mu$m~$\times$~$20$~$\mu$m area in a meander configuration \cite{Miki2008}.  Two such SNSPD chips are fiber coupled and mounted inside a closed cycle refrigerator at a temperature of approximately $2.4$~K \cite{Hadfield2005}.  In this study we focus in detail on the behavior of one of the fiber-coupled SNSPDs.

\begin{figure} [!h]
\center
\resizebox{8cm}{!} {
\includegraphics{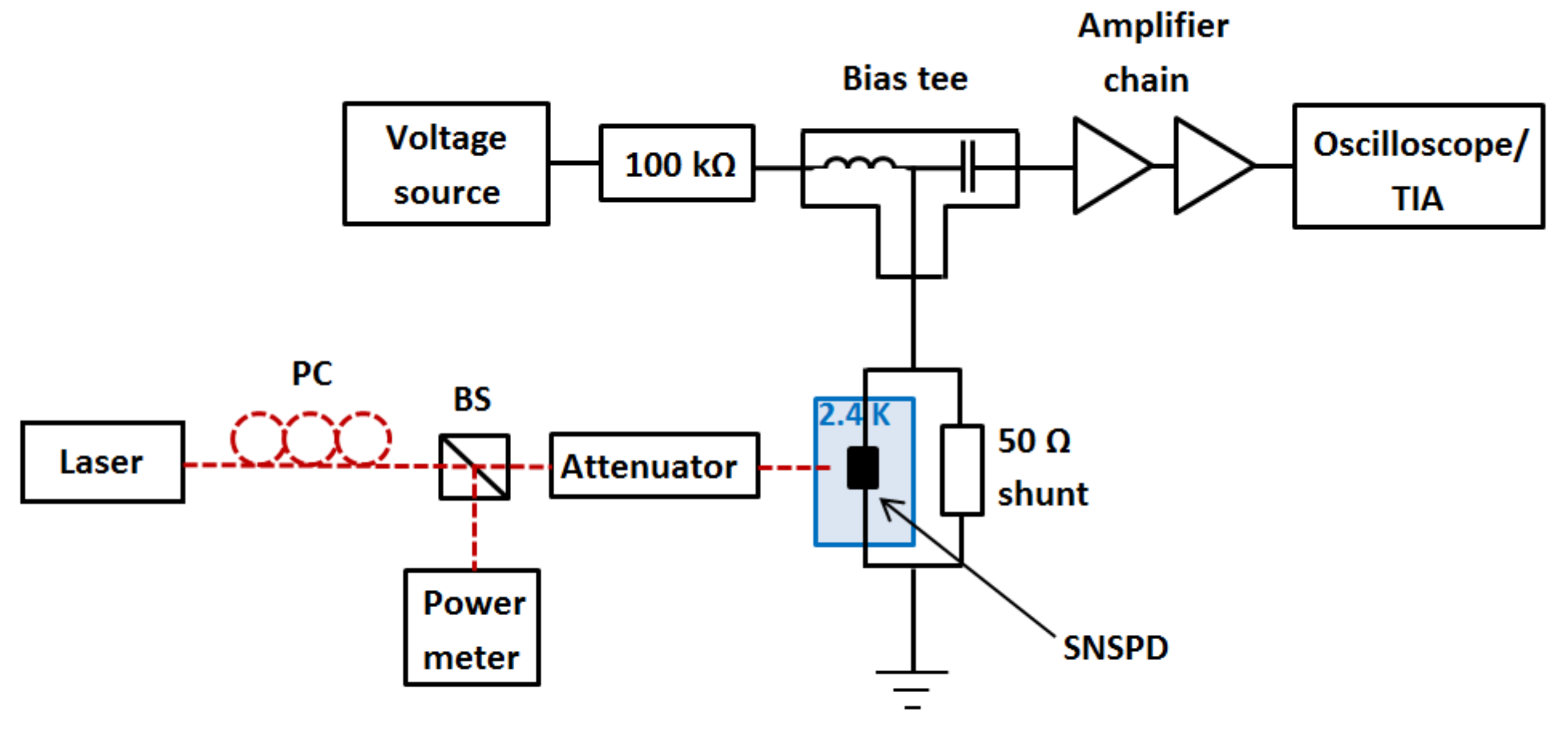} }
\caption{Schematic set-up of the Superconducting Nanowire Single-Photon Detector (SNSPD) and associated components. Dashed red lines represent optical fiber; solid black lines represent electrical connections. An attenuated polarization-controlled laser provides the optical input to the detector chip. A battery powered voltage source and a $100$~k$\Omega$ resistor is used to bias the nanowire just below its critical current through one branch of the bias tee. The output electrical signal from the SNSPD is read out using the other branch of the bias tee, and is amplified by the amplifier chain before being detected. The SNSPD is maintained at a stable temperature of $2.4$~K inside a closed-cycle refrigerator (Sumitomo RDK101D cold head and CNA11C compressor). The length of the cable between the SNSPD and tee piece for the $50$~$\Omega$ shunt, and between the SNSPD and the bias tee is about $1.5$~m of coaxial cable, almost all of which is inside the cryocooler. PC: polarization controller, BS: beam-splitter, TIA: Time Interval Analyzer.}
\label{fig: setup}
\end{figure}

The schematic of our SNSPD set-up is shown in Fig.~\ref{fig: setup}. It is a fairly standard configuration.

The laser used was a pulsed laser (PicoQuant PDL 800-B) at $1550$~nm. An optical attenuator together with a power meter is used to set the desired power of incoming light. The polarization controller (PC) is used to adjust the polarization of incoming light to maximize the detection efficiency \cite{Anant2008}.

The SNSPD is current-biased below the critical current by setting a DC bias on the battery-powered voltage source. A bias tee (Picosecond Pulse Labs, part-ID 5575A-104, $12$~GHz bandwidth), which is essentially a combination of an inductor and a capacitor, is used to set the DC bias through one arm and read out the RF signal through the other arm.

The weak electrical output signal from the detector is then amplified as it passes through the amplifier chain. The amplifier chain consists of two amplifiers, RF-Bay LNA-580 and LNA-1000, with a combined gain of around $56$~dB. The LNA-580 and LNA-1000 amplifiers have a $3$~dB roll off of $580$~MHz and $1$~GHz respectively.

The amplified signal is then detected by a Time Interval Analyzer (TIA), which records the arrival time of detector output signals and synchronization signal with picosecond resolution. The TIA model is PicoQuant HydraHarp 400. The TIA dead time per channel is specified as $<$80~ns. Thus the TIA dead time plays a negligible role in our investigations since it is considerably lower than the dead time of our SNSPD, which is of the order of $150$~ns \footnote{See supplementary material at [URL will be inserted by AIP] for a discussion about the dead time of our SNSPD.}.

Note that we are using a $50$~$\Omega$ shunt resistor in our SNSPD set-up to avoid SNSPD latching \cite{Hadfield2005a}. Latching is a phenomenon which can occur in SNSPDs where the nanowire doesn't recover to its superconducting state after a detection \cite{Kerman2009}. The use of the shunt resistor doubles the dead time of our SNSPD but was necessary in our set-up to avoid latching in the high bias regime we were working in \footnote{See supplementary material at [URL will be inserted by AIP] for more details about the latching phenomenon.}.

The SNSPD bias can be set at different values by changing the voltage of the voltage source depending on what the SNSPD is being used for. For this specific type of NbN SNSPD, higher bias gives a higher detection efficiency, as shown in \cite{Natarajan2012} Fig 3(a). In other types of SNSPDs, the ‘plateau’ behavior is observed where efficiency saturates at high bias \cite{Marsili2011, Baek2011}.  This can be observed in very uniform short NbN nanowires at short wavelengths, or in new materials like WSi at telecom wavelengths.

Higher bias in our SNSPD also raises the dark count rate (DCR) together with the detection efficiency. Applications in which it is essential to minimize dark counts, such as long distance QKD \cite{Hadfield2006, Takesue2007, Stucki2009}, would require the SNSPD to be operated at a relatively low bias. Above the critical current, the SNSPD will undergo relaxation oscillations (outputting a continuous train of pulses). The maximum detection efficiency of the SNSPD is approximately 2.5$\%$, at a bias of about $25$~$\mu$A.

\begin{figure} [!h]
\center
\resizebox{7cm}{!} {
\includegraphics{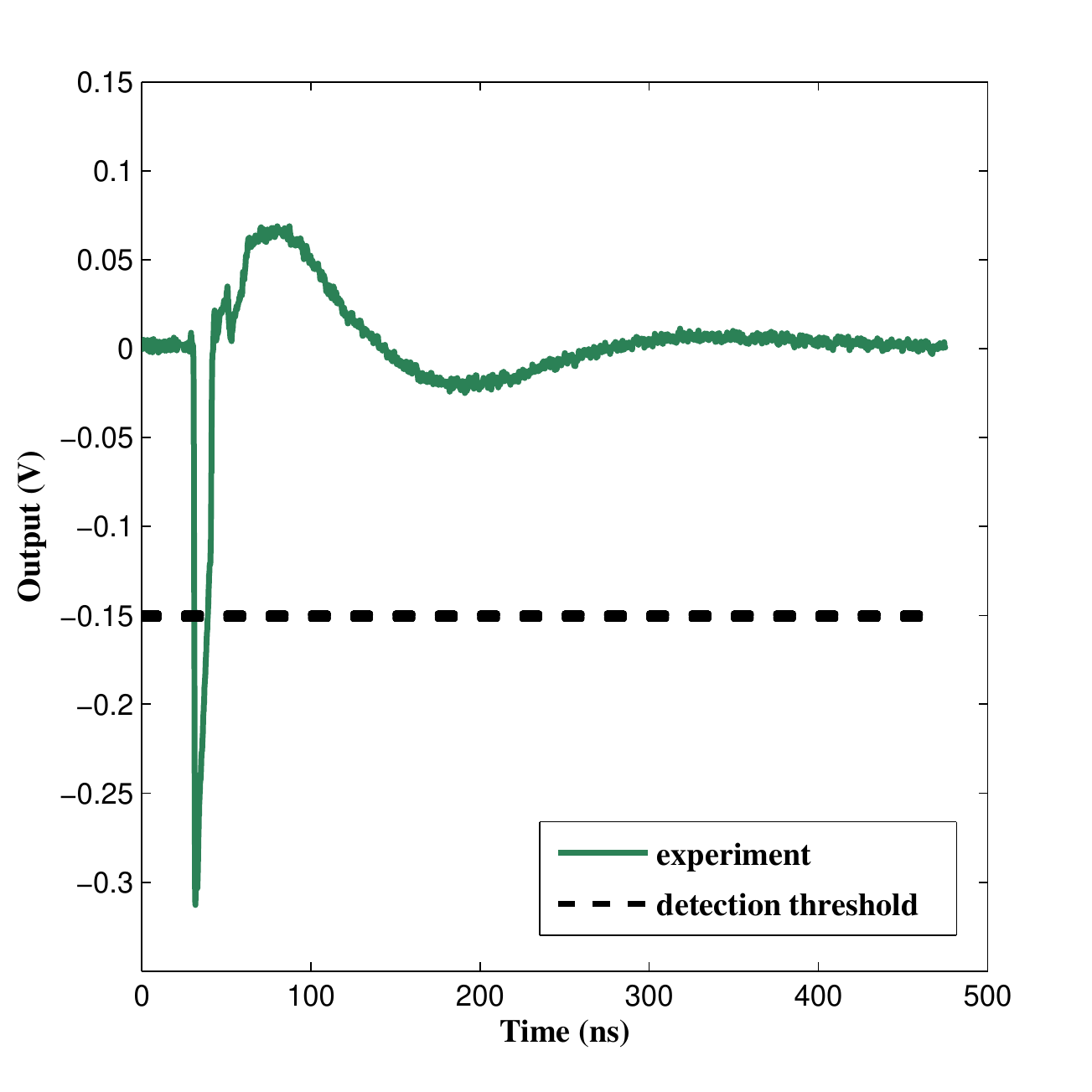} }
\caption{Output pulse shape of a single detection click, starting at $30$~ns, as observed on an oscilloscope (rugged green line). The observed shape shown is the average of ten pulse shapes to reduce appearance of random noise, but the individual pulse shapes are nearly identical to each other. The dotted black line shows the discrimination voltage level used by the TIA to register a detection event.}
\label{fig: pulseThreshold}
\end{figure}

It is important to define what exactly constitutes a single detection event. The output voltage pulse coming from the amplifiers of a single detection click, over a time span of $500$~ns, is shown in Fig.~\ref{fig: pulseThreshold}.

Fig.~\ref{fig: pulseThreshold} shows a detection event which occurs at $30$~ns, causing a rapid drop of around $300$~mV, followed by an overshoot and further oscillations in the voltage pulse. The shape and height of the pulses is very consistent between different detection events (the height of the peak (dip) is proportional to the bias current of SNSPD). We use the leading edge of the pulse to discriminate a detection event with the TIA, setting the discrimination level at negative $150$~mV. This is roughly halfway down the pulse and this amplitude value is well beyond the noise level.

\section{Afterpulsing of Dark Counts}
\label{Section4}

In this section, we study the dark count distribution of the SNSPD. Therefore, we turn the laser off. We shall focus on results with the bias current set to $25.0$~$\mu$A. This bias is close to the critical value and hence maximizes the detection efficiency of the SNSPD, albeit at the cost of a higher DCR.

The mechanisms for dark counts in SNSPDs have been discussed in Ref. \cite{Natarajan2012, Bulaevskii2011} and references therein. Even if care is taken to minimize stray light and blackbody radiation contributions to the dark count, there would still be a finite amount of dark clicks. The dominant mechanism for this is believed to be current-assisted unbinding of vortex-antivortex pairs \cite{Yamashita2011}, although other mechanisms have also been proposed.

It is generally assumed and rarely questioned that dark counts occur randomly and uniformly in time. The waiting times between each pair of consecutive dark count events (inter-arrival time) are independent random variables, and it is thus a Poisson process. As such these inter-arrival times should follow an exponential distribution. We can see this intuitively as follows. For each fixed-time interval the probability of having a dark count is the same. However, after some starting point in time the probability of having a dark count in each successive fixed-time interval decreases (exponentially) since for a click to happen further down the timeline implies a click did not happen in all prior fixed-time intervals. The single likeliest fixed-time interval where a click would happen is the first one, followed by the second one, and so on.

We decided to verify this by plotting a histogram of waiting times between dark count events. In this experiment, the laser is switched off and we are measuring dark count events only. The detection times are recorded with the TIA. We calculate time difference between neighboring clicks, and group these into $0.1$~ms time windows ($=$~bin size) to plot the histogram.

\begin{figure} [!h]
\center
\resizebox{8cm}{!} {
\includegraphics{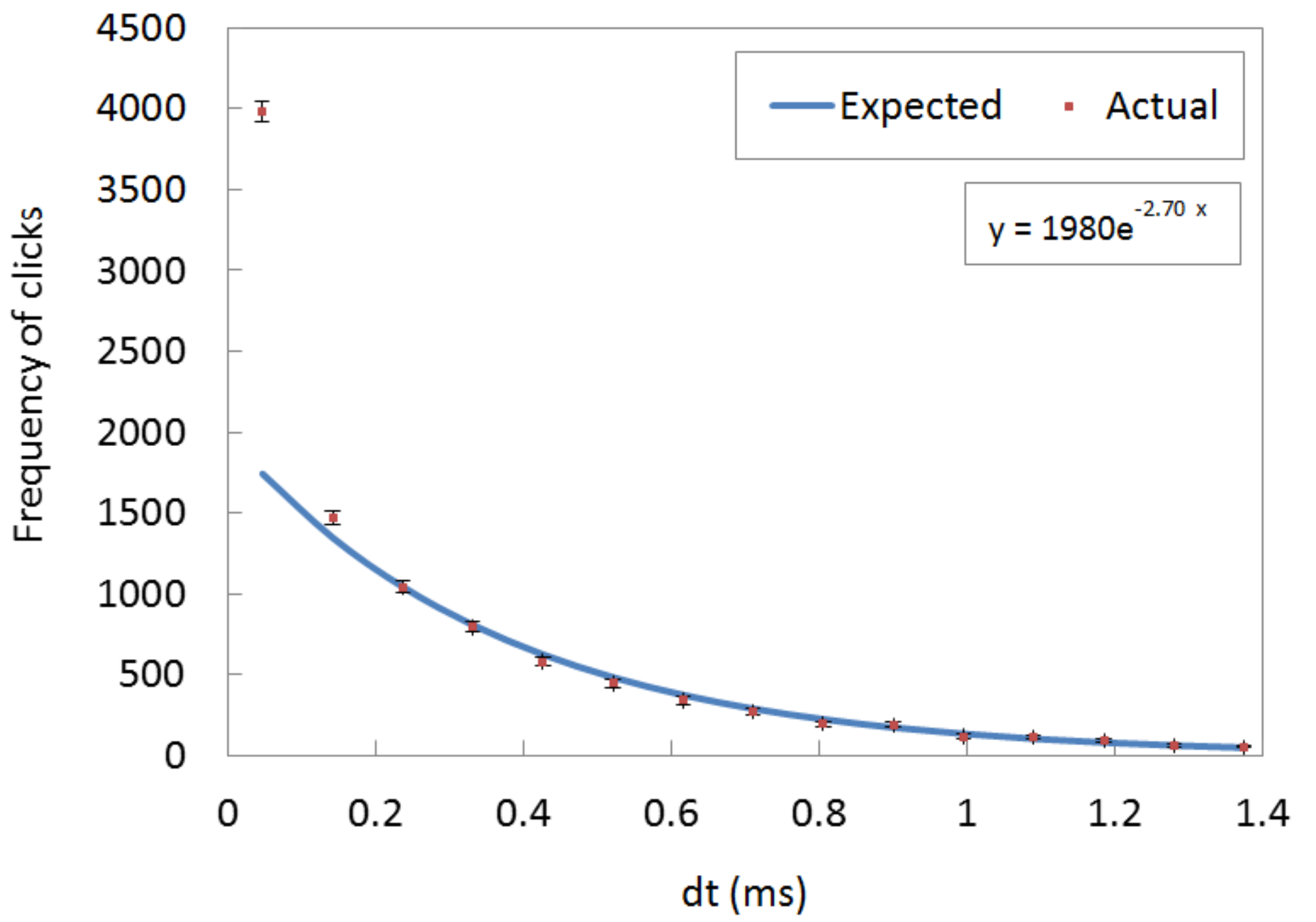} }
\caption{Histogram of waiting time between dark count clicks. An exponential decay fit (blue line) is superimposed on the data points (red dots). The acquisition time was approximately $3$~s giving 10,000 detection events in total. The error bars shown are given by the square root of the number of detection events. The first 15 bins are shown. $I_b = 25.0$~$\mu$A. $I_c = 25.3$~$\mu$A.}
\label{fig: darkexp}
\end{figure}

The bias was set to a relatively high value of $25.0$~$\mu$A, which is near the critical value. The DCR at this bias level is approximately $3200$~counts/s. Histogram of waiting times dt between dark counts is shown in Fig.~\ref{fig: darkexp}.

The only expected change from this theoretical model would be due to the detector's dead time. However, since the dead time of our detector is of the order of $150$~ns, this effect would be negligible with relatively large bin size of the order of $0.1$~ms.

We can see from the graph that the data points (bin heights at center of bin) seem to follow an exponential decay except for the first bin, which clearly has a larger than expected value. We fitted an exponential decay curve on top of the data points. To fit the exponential, we discarded the first bin, and also all bins towards the tail end of the distribution (with small values in each bin). We then extrapolated the given curve (blue line in Fig.~\ref{fig: darkexp}) to all bins. There is good agreement amongst all the other bins except for the first bin.

\begin{figure} [!h]
\center
\resizebox{8cm}{!} {
\includegraphics{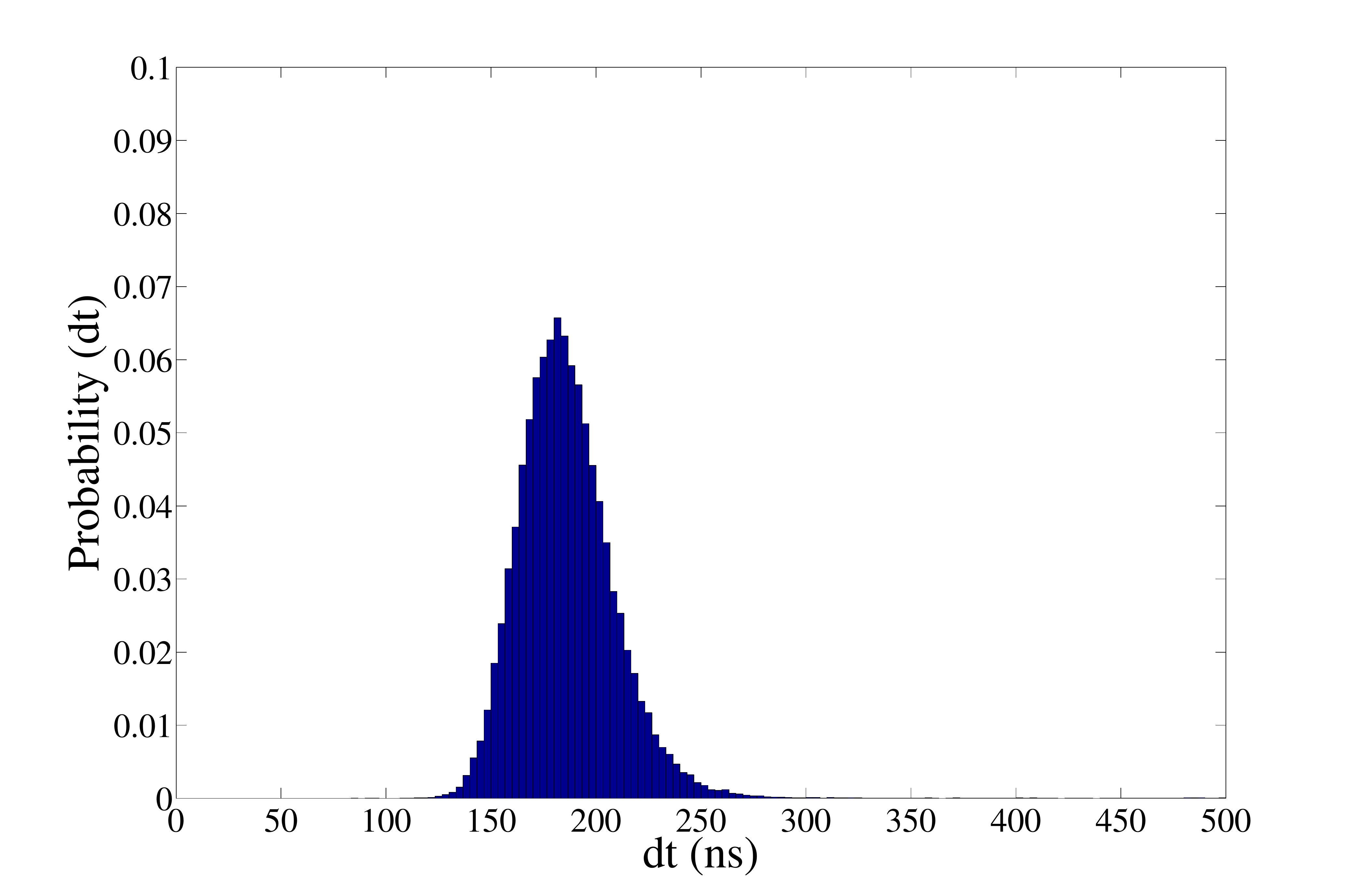} }
\caption{Histogram of waiting time between dark count clicks, for the first $500$~ns after an initial click. Bias current $= 25$~$\mu$A. 100, 000 events total, bin width $= 4$~ns. $I_b = 25.0$~$\mu$A. $I_c = 25.3$~$\mu$A.}
\label{fig: darkpeak}
\end{figure}

Zooming into the first bin clearly reveals the presence of an afterpulsing effect. We plotted a histogram of waiting times within the first $500$~ns using a much smaller bin size ($4$~ns). This is a separate experiment with a longer acquisition time than in Fig.~\ref{fig: darkexp}. The histogram for the first $500$~ns is shown in Fig.~\ref{fig: darkpeak}.

The figure shows a large number of unexpected counts in an approximately Gaussian distribution centered at around $180$~ns waiting time. That is, there is an increased probability of a dark count effect occurring around the $180$~ns mark after a previous dark count. This is why we refer to this effect as afterpulsing. This effect was independently verified with an oscilloscope, and so it cannot be an artifact of our TIA system.

We can define the amount of afterpulsing as the number of clicks in the first $1000$~ns window (there would be negligible real dark counts in such a short time interval). Since the afterpulse effect predominantly happens within the first $1000$~ns of a previous click, it is useful to introduce a new quantity called \textit{corrected} DCR, which is the total DCR minus all the clicks within the first $1000$~ns, i.e. the afterpulses.

\begin{figure} [!h]
\center
\resizebox{8cm}{!} {
\includegraphics{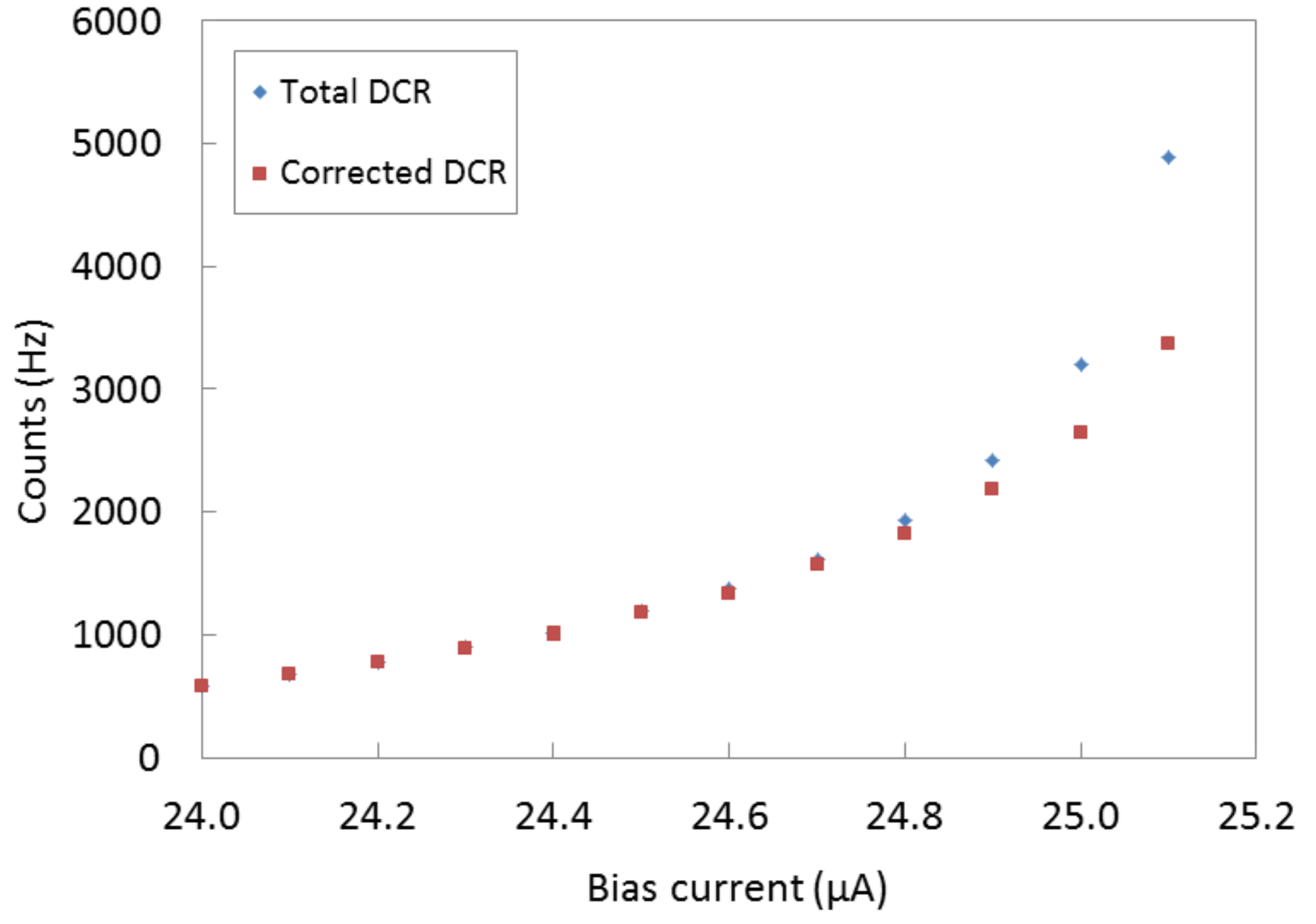} }
\caption{Total DCR and corrected DCR versus bias current. About 10,000 total clicks were acquired for each bias point. At higher bias, the two deviate more due to more prevalent afterpulsing. The error bars for each point are smaller than the size of the marker. $I_c = 25.3$~$\mu$A.}
\label{fig: correctedDCR}
\end{figure}

A plot of how total DCR and corrected DCR vary with bias current is shown in Fig.~\ref{fig: correctedDCR}.

We can see that at lower bias value the total DCR closely matches the corrected dark count rate. At higher bias values the two deviate more and more, as the afterpulsing contributes more to the DCR.

\begin{figure} [!h]
\center
\resizebox{8cm}{!} {
\includegraphics{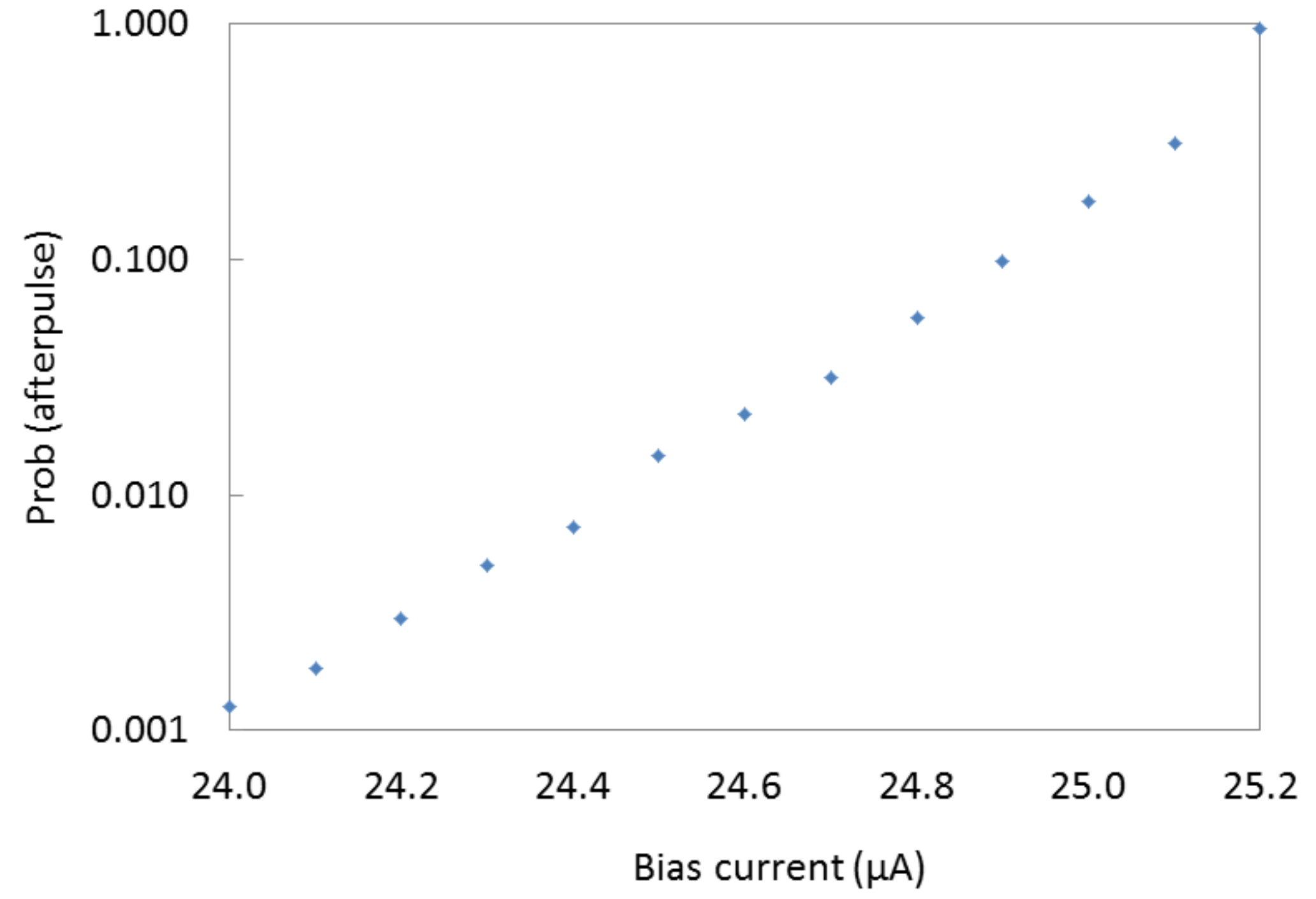} }
\caption{Afterpulse probability vs bias current. The afterpulsing effect quickly becomes negligible at the bias is reduced away from the very high value of $25.2$~$\mu$A. $I_c$ is about $25.3$~$\mu$A.}
\label{fig: probafterpulsing}
\end{figure}

We now examine how changing the bias current affects afterpulsing. We define the probability of afterpulsing as the number of clicks that happen within $1000$~ns of a prior click divided by the total number of clicks. We discovered that the afterpulsing strongly depends on the bias current. A plot of afterpulse probability versus current bias is shown in Fig.~\ref{fig: probafterpulsing}.

We see an exponential increase in the afterpulsing probability as the bias current approaches the critical value. Afterpulsing quickly becomes negligible as bias is decreased away from the critical value.

We noticed that afterpulses can occur in trains of one or more afterpulses. Whether the second afterpulse, typically occurring about $360$~ns after the initial click, is simply the afterpulse of the first afterpulse is investigated in Sec.~\ref{Section5}.

\section{Model for afterpulsing}
\label{Section5}

\begin{figure} [!h]
\center
\resizebox{9.5cm}{!} {
\includegraphics{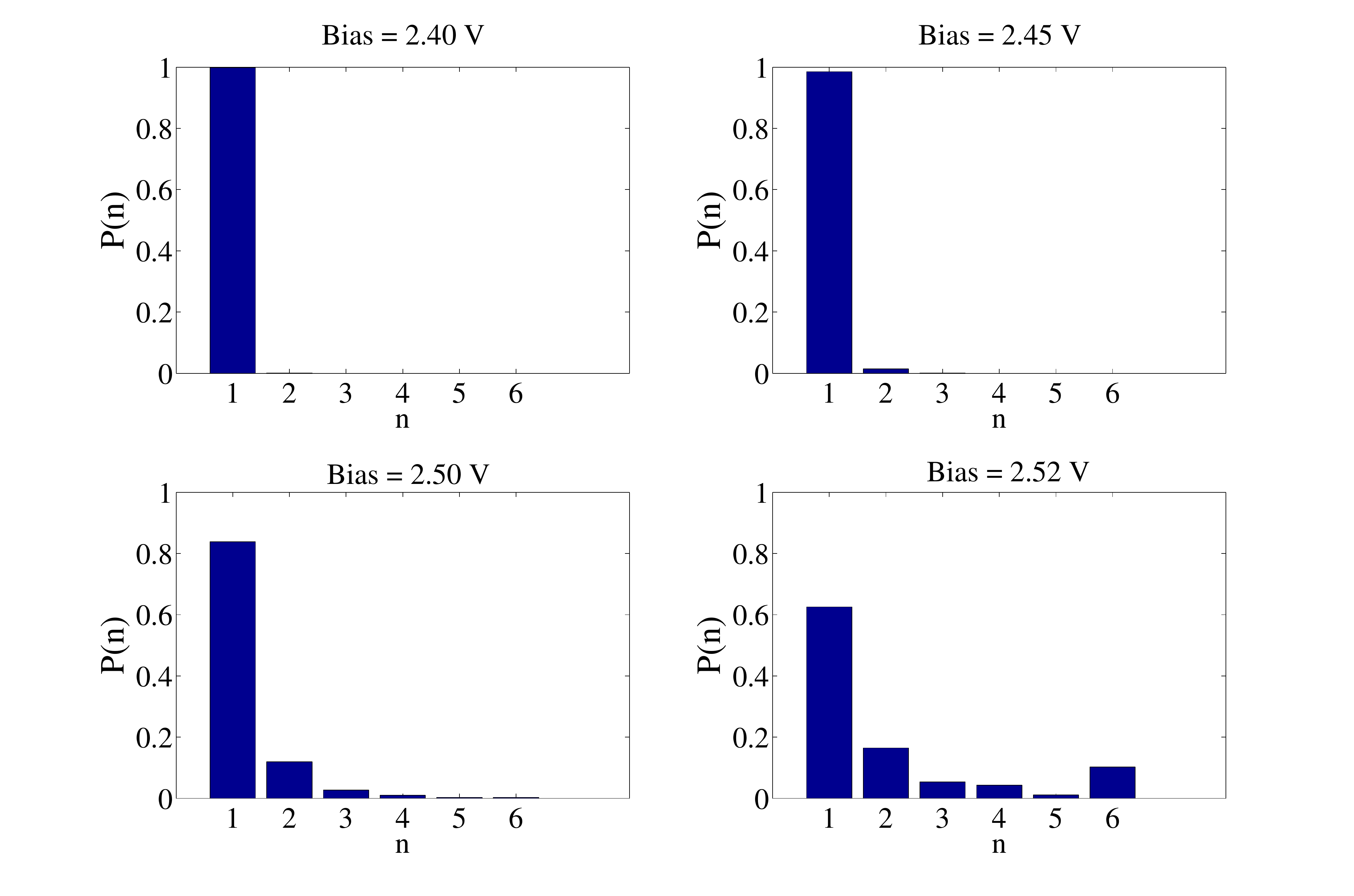} }
\caption{This figure shows how the observed afterpulses start to appear in multiples of clicks as the bias is increased towards the critical value, of about $25.3$~$\mu$A. The $n = 6$ column represents "6 or more" pulses.}
\label{fig: pngraphs}
\end{figure}

It is important to characterize and understand the underlying cause of the afterpulsing in SNSPDs. For this we have measured the number of afterpulses that occur in the afterpulse `train' of n clicks, at different bias values, with no light input.
A single click corresponds to $n = 1$. A two-pulse train is $n = 2$, i.e. a single afterpulse click (occurring at around $180$~ns), a three-pulse train is $n = 3$ (first afterpulse at approximately $180$~ns, second at approximately $360$~ns), and so on. We define the trains such that a train with a certain number of pulses in it is distinct so that, for example, an n-click train does not also count as an (n-1)-click train.
The afterpulse trains are detected using the TIA, which records all the dark counts over a period of time, and the output is processed on a computer to compile a histogram to show the number distribution of clicks in these afterpulse trains. See Fig.~\ref{fig: pngraphs}.

\begin{figure} [!h]
\center
\resizebox{8cm}{!} {
\includegraphics{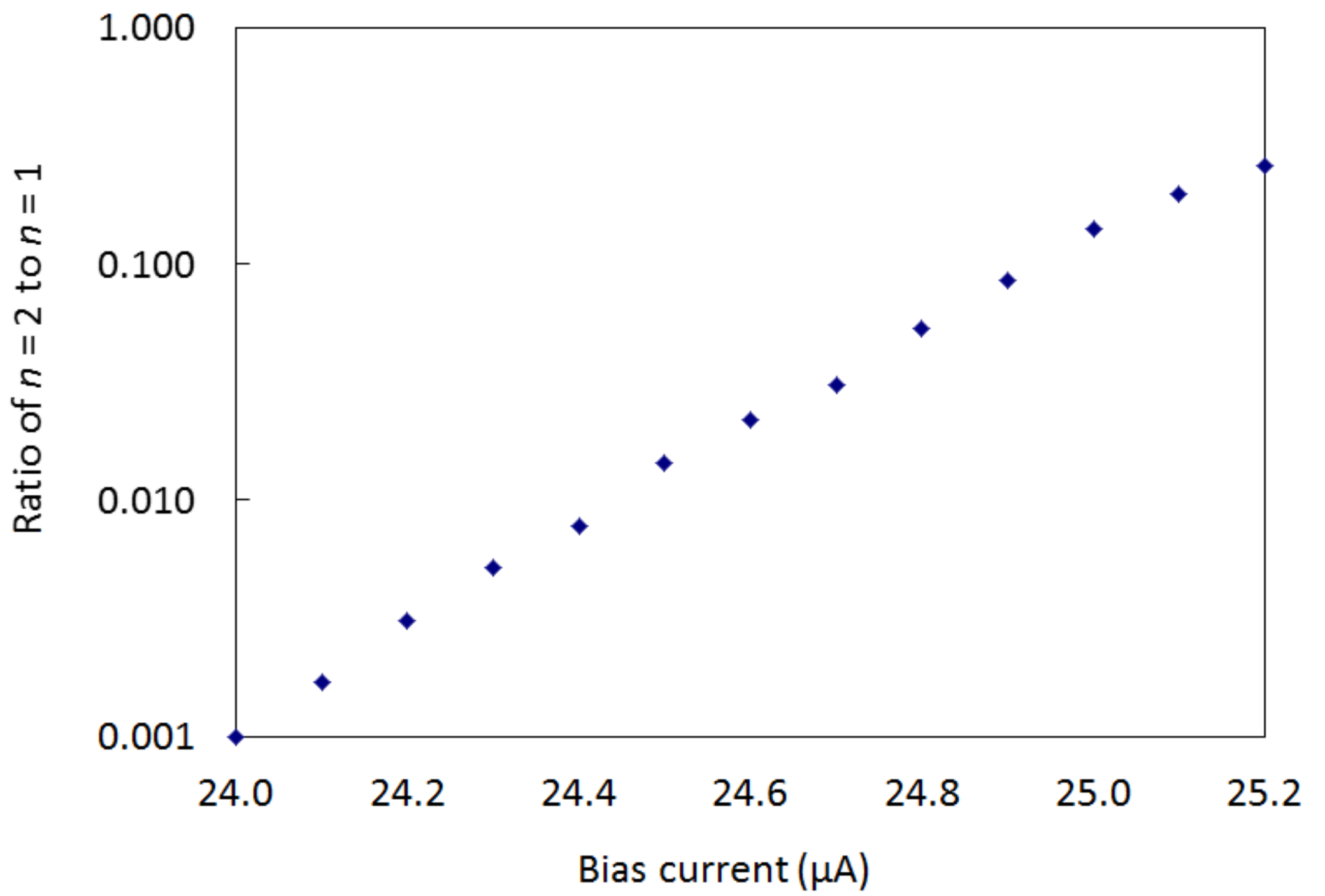} }
\caption{Ratio of $n = 2$ (two clicks with the second being an afterpulse click) to $n = 1$ (single click) in $P(n)$ vs $n$ graphs of Fig.~\ref{fig: pngraphs}, at different bias values. $I_c = 25.3$~$\mu$A.}
\label{fig: n1n2}
\end{figure}

We can see that at lower bias there are virtually no afterpulses. They start to appear as single afterpulses, but as the bias is increasing they start to come in pairs, triplets and more afterpulses in one afterpulse train. For our purposes it is useful to look at the ratio of $n = 2$ (i.e. single afterpulse click) to $n = 1$ events (regular dark count), which is shown in Fig.~\ref{fig: n1n2}.

Fig.~\ref{fig: n1n2} shows an exponential increase in the ratio as the bias current approaches the critical value. Moreover, it gives the same slope as in Fig.~\ref{fig: probafterpulsing}.

We can see that there seems to be a simple model to describe the secondary afterpulses for a large range of bias values, although the model breaks down when the bias level gets very close to critical. The model being that subsequent afterpulses in the train are caused by previous afterpulses.

In agreement with earlier work in Ref. \cite{Fujiwara2011} we believe that the afterpulses are caused by reflections in the readout circuit.  The transient of the voltage pulse may perturb the grounding of the circuit, perturbing the bias current of the SNSPD. This has very little effect if the operational bias is well below the critical value. However, if the operating value is close to critical, this small increase in the effective bias voltage is enough to make a big difference in both the detection efficiency and the dark count rate for a short period of time around $180$~ns after the initial detection. This also explains how the bias current can change the afterpulsing probability. Given these results, we have no reason to think that the actual SNSPD itself has intrinsic afterpulsing.

\section{Afterpulsing with laser on}
\label{Section6}

In this section we use a pulsed laser running at $0.5$~MHz to investigate the afterpulse phenomenon for actual detection events. Note that the experiments from this section onwards were taken on a different day, and the temperature of the SNSPD had increased slightly so that the bias current was lowered slightly from about $25.0$~$\mu$A to $24.5$~$\mu$A to approximately match the biasing level of the SNSPD making it consistent with previous results. The critical current had decreased from about $25.3$~$\mu$A to about $24.8$~$\mu$A. We used the TIA to obtain a record of all the detection events. The laser provides a synchronization signal to the TIA.

\begin{figure} [!h]
\center
\resizebox{8cm}{!} {
\includegraphics{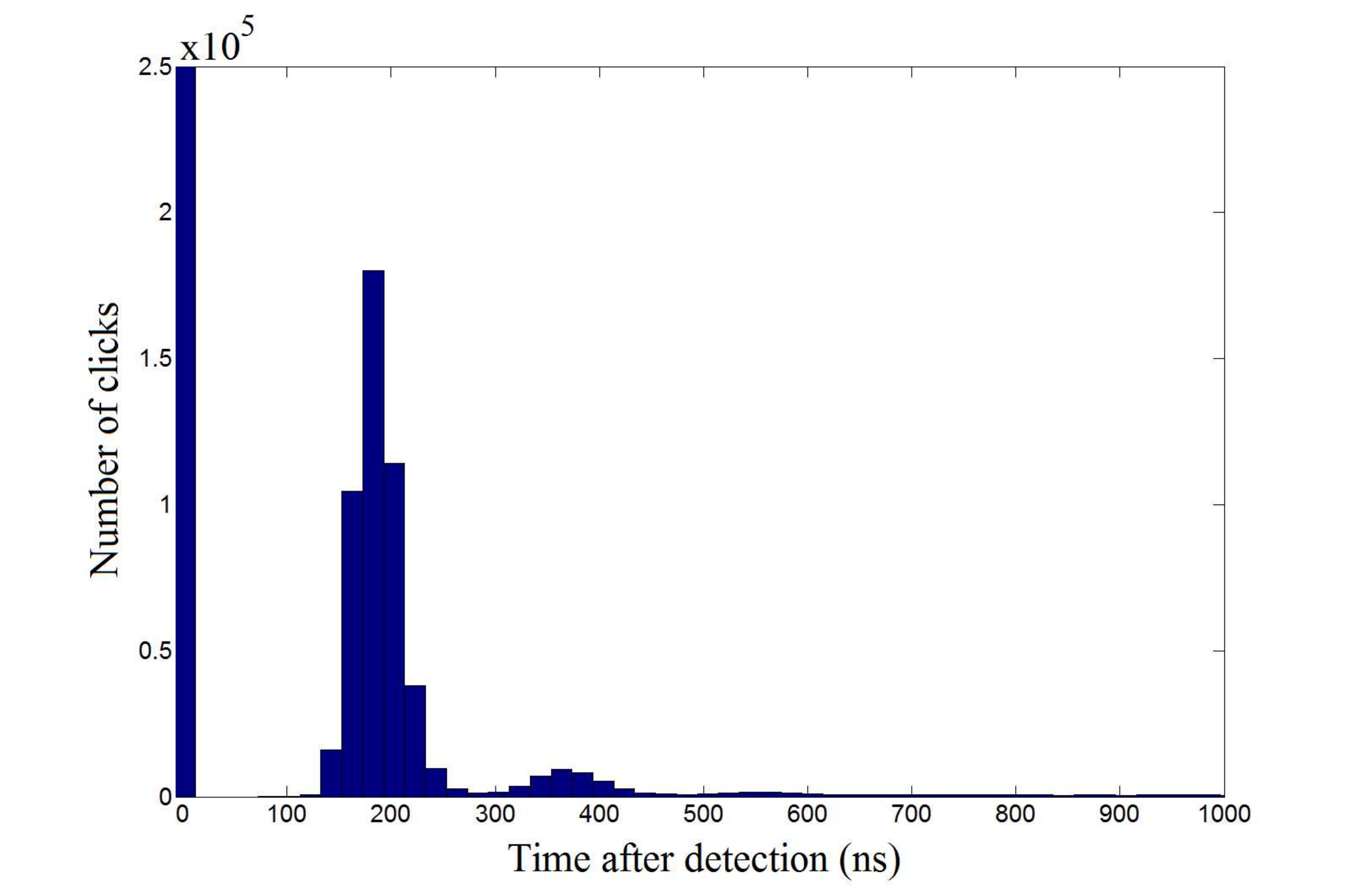} }
\caption{Histogram of time interval between the laser pulse signal and detection click. Number of total clicks is just over $17$~million, almost all of which are concentrated in the first peak at $t = 0$, which corresponds to the detection time of the laser pulses. Laser repetition rate is $0.5$~MHz. Photon number $\mu = 10$ photons/pulse. Bin size is $20$~ns. $I_c = 24.8$~$\mu$A.}
\label{fig: laseron}
\end{figure}

We run the experiment continuously for $150$~seconds. Using the sync signal as time reference, we split up the timeline of detection events into $2000$~ns segments. We then record the time interval between the click caused by the laser pulse and any other clicks in that $2000$~ns time window. We build up a histogram of these time intervals over all $2000$~ns segments. The graph is adjusted in the time domain so that the arrival of the electrical signal from the SNSPD caused by the laser pulse, is at $t = 0$. We also only count clicks in one of these $2000$~ns segments if the laser pulse did in fact cause a corresponding detection event in that segment. In other words, we are counting further detection clicks conditionally in the event that there was a click at $t = 0$ caused by the laser pulse. This histogram of time interval between the laser pulse signal and detection clicks is shown in Fig.~\ref{fig: laseron}.

The tall sharp peak at $t = 0$ corresponds to the position of the laser pulse. The other broader peaks correspond to afterpulses. The first broad peak appears at around $180$~ns. This is consistent with the results with no laser input. We also see the SNSPD dead time at the beginning. The baseline level corresponds to regular dark counts.

We confirm that the aforementioned afterpulsing effect is present for the case of pure dark count clicks (no laser input) as well as real detection clicks (from pulsed laser), and the peak of the afterpulse occurs after the same delay (about $180$~ns) in both cases.

\section{Detection efficiency recovery}
\label{Section7}

One might ask whether the afterpulse peak around $180$~ns is caused by a higher detection efficiency in the SNSPD at this time, or if it is simply a result of probabilistic events that either happen or not, regardless of detection efficiency (which could in principle, based on results presented so far, even be zero). In this section we answer this question and show that indeed the detection efficiency around that point is higher. We also show how the actual detection efficiency recovers back to its nominal value following a detection. The recovery is not monotonic, and in fact the detection efficiency oscillates in the first $500$~ns before settling to the expected value. A measurement of the actual recovery process was reported in Ref. \cite{Kerman2006}, giving a smooth recovery to the nominal value.

To measure detection efficiency dynamics, we perform the following `double pulse' experiment. The pulsed laser is used to make pairs of pulses in $2000$~ns windows, at variable separation, ranging from $80$~ns to $1000$~ns. These pairs of pulses are sent into the SNSPD and the sync signal is matched to the first laser pulse of the pair. The number of photons in each laser pulse is $\mu = 1$.

\begin{figure} [!h]
\center
\resizebox{8cm}{!} {
\includegraphics{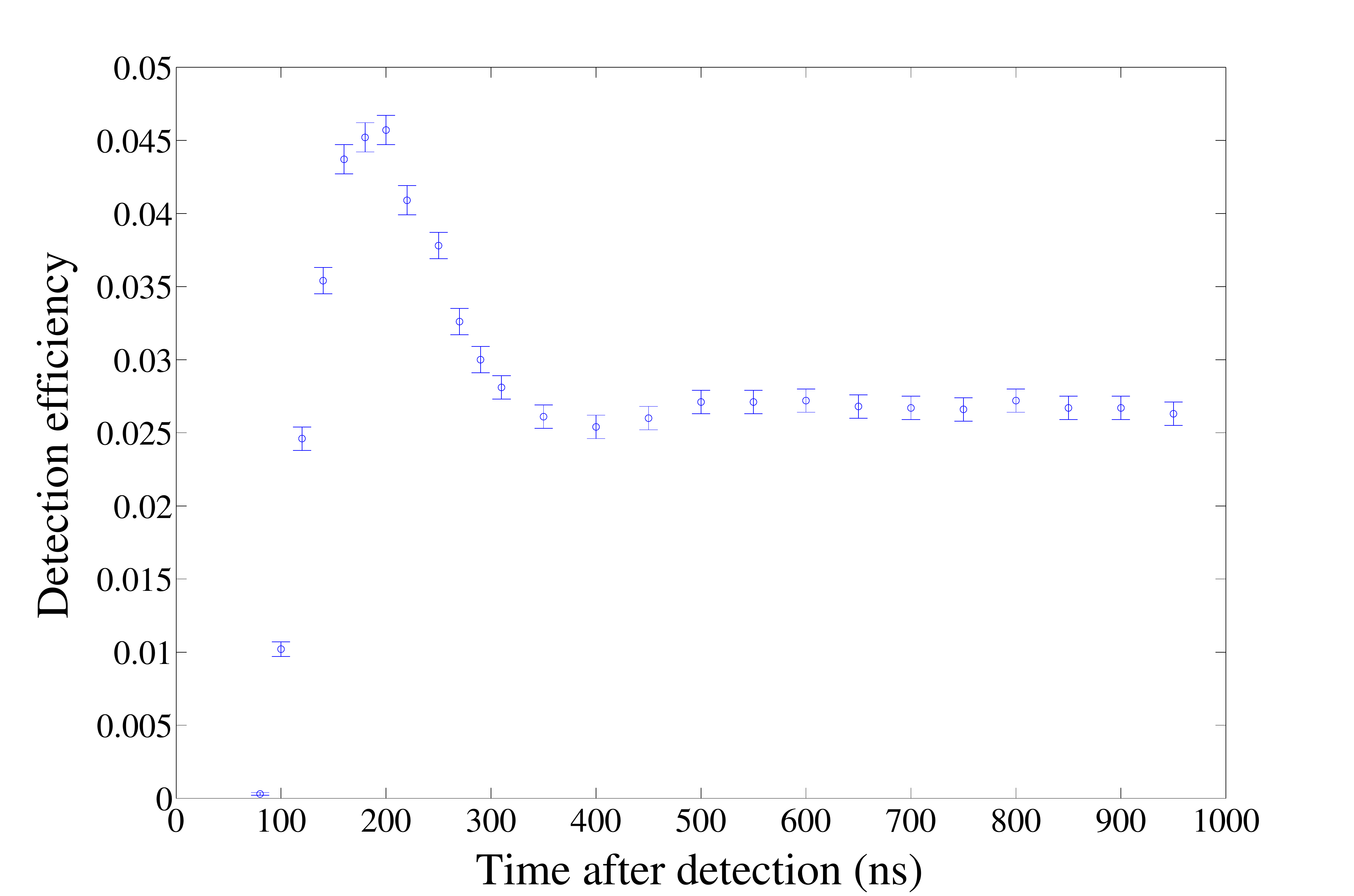} }
\caption{Detection efficiency recovery graph of our SNSPD. The delay between the two pulses ranges from $80$~ns to $1000$~ns. Each point is obtained by setting the TIA acquisition time to $30$~seconds, resulting in about 400,000 instances of the first laser pulse being detected. The detection efficiency recovery is not monotonic. The error bars for each point only take into account the statistical fluctuations, defined by $\pm 3\sqrt{number~of~events}$. The bias point of the SNSPD was $24.4$~$\mu$A (the critical current at the time was about $24.8$~$\mu$A).}
\label{fig: recovery}
\end{figure}

To calculate the detection efficiency for the second pulse in the pair, we take the ratio of the number of cases where \textit{both} pulses were detected to the number of cases where only the first pulse was detected. We subtract the estimated number of afterpulse clicks caused by the first pulse in the pair \footnote{See supplementary material at [URL will be inserted by AIP] for details of the procedure used to make these estimates.}. The resulting graph of detection efficiency recovery is shown in Fig.~\ref{fig: recovery}.

The result is unexpected as the increase to nominal detection efficiency is not monotonic. The detection efficiency is virtually zero up to about $80$~ns, at which point it starts to rise sharply, reaching a peak at around $180$~ns after the detection, corresponding to the same point in time where afterpulsing is at its highest. This is followed by a drop to the expected nominal level after several hundred nanoseconds after a detection.

Based on this evidence, it seems plausible to us that both the afterpulsing and the unexpected detection efficiency recovery are caused by the same external phenomenon. We speculate that the transient in the voltage pulse is perturbing the current flow in the nanowire, leading temporarily to higher bias current, and thus an enhancement of detection efficiency around $180$~ns after a detection event. Likewise, we speculate that this temporary overshoot in the bias current could lead to a temporary increase in dark count rate around $180$~ns after a detection event, thus causing the observed afterpulsing. However, more experiments would need to be done to make a definitive conclusion.

It is worth noting that the overshoot of the response pulse and the detection efficiency reach their maxima at different times after the falling edge of the pulse crosses the threshold. One explanation for this mismatch could be that the temporal voltage pulse profile recorded on the oscilloscope gives only an indirect indication of the current recovery in the nanowire.  There may be a lag between the two owing to the impedance of the circuit.

We replace the amplifier chain with an alternative amplifier and show that the afterpulsing disappears (see Sec.~\ref{Section8}).

\section{Using an amplifier with improved low frequency response}
\label{Section8}

\begin{figure} [!h]
\center
\resizebox{8cm}{!} {
\includegraphics{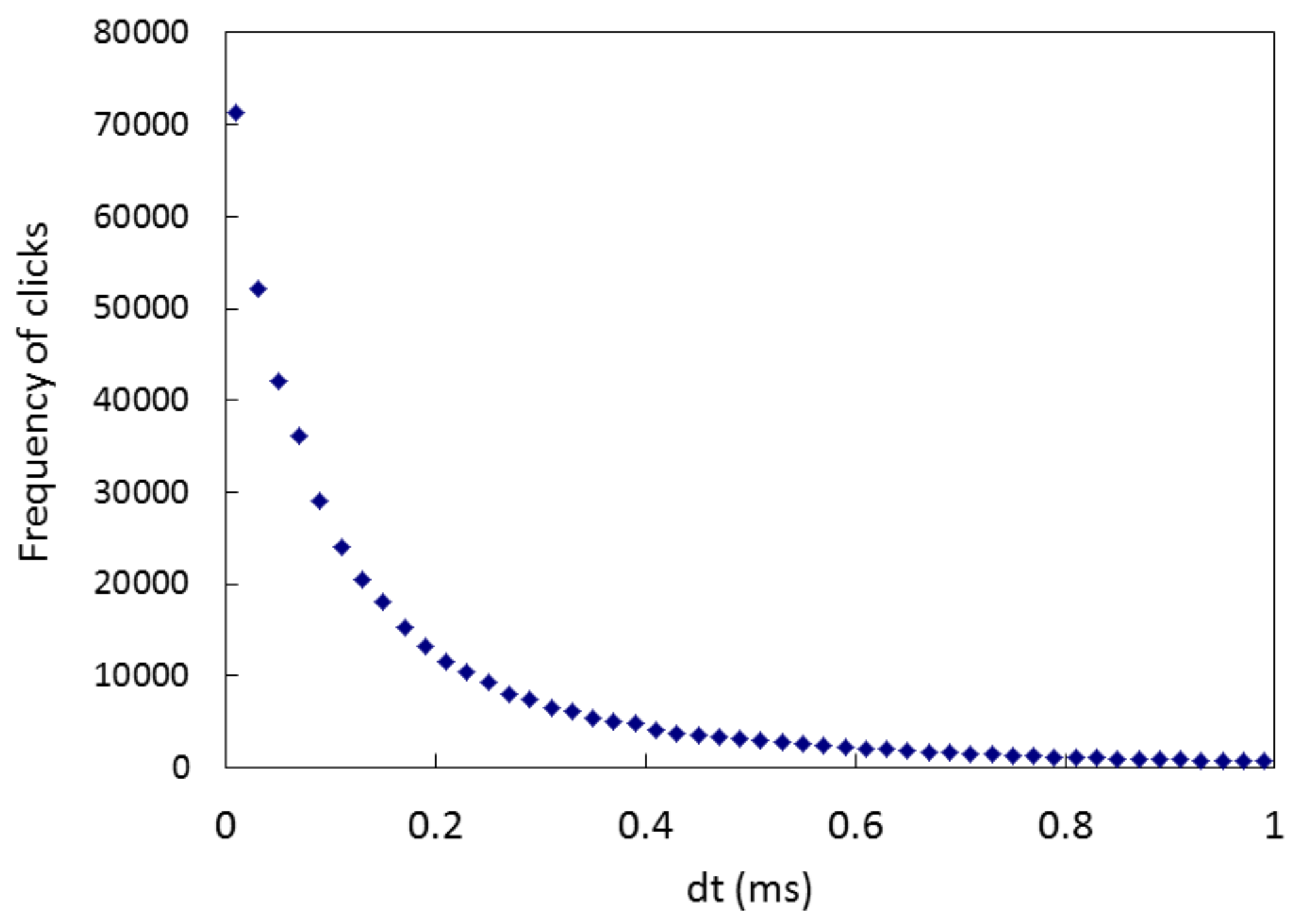} }
\caption{Histogram of waiting time between dark count clicks. There are approximately 450,000 events in total. The bin size is $20,000$~ns. $I_c = 24.8$~$\mu$A. Afterpulsing is no longer present.}
\label{fig: darkexp2}
\end{figure}

We replaced the two RF Bay amplifiers in our set-up with a single MITEQ AM-1431 amplifier, which has a frequency range of $0.001$~--~$1000$~MHz. This amplifier was chosen due to its improved low frequency response compared to the RF Bay amplifiers, which have frequency ranges of $10$~--~$580$~MHz and $10$~--~$1000$~MHz. We repeat the experiment of acquiring dark counts from Sec.~\ref{Section4}, and plot the histogram of waiting times between dark count events. See Fig.~\ref{fig: darkexp2}.

Although we did not get a chance to optimize the conditions to directly compare the new result to our previous result from Sec.~\ref{Section4} (due to technical issues with our SNSPD cooling system), we have observed that there is no longer an afterpulsing peak within the first $500$~ns after a dark count, and as such, there is no longer an abnormally high number of clicks in the first bin as we can see from Fig.~\ref{fig: darkexp2}.

We speculate that the poor low frequency response of the RF Bay LNA-580 and LNA-1000 amplifiers used caused an oscillation in the output voltage pulse during recovery, leading to the afterpulsing effect, as suggested in Ref. \cite{Fujiwara2011}.

\section{Summary and Conclusions}
\label{Section9}

The contributions of this paper can be summarized in terms of three main parts. First, we have performed the first experimental study of the time-domain distribution of dark counts in an SNSPD (i.e. the case with no laser input), showing for the first time, an afterpulsing effect. Second, we have also performed a longer time domain measurement of the afterpulsing effect ($1000$~ns) with laser input. Third, we have measured the actual recovery graph of the detection efficiency and found that the recovery occurs in an unexpected manner. Specifically, we found that the detection efficiency after a detection event starts from zero, rises and actually overshoots before returning to its normal value.

The afterpulsing, dark count and detection efficiency enhancement during the recovery of the SNSPD show similar dynamics and are all most likely to occur around $180$~ns following a detection event (the exact figure depends on the biasing current). We speculate that all three of these effects arise due to a perturbation in the bias current through the nanowire. This is caused by the poor low frequency response of the RF Bay LNA-580 and LNA-1000 amplifiers used, causing an oscillation in the output voltage pulse during recovery, as suggested in Ref. \cite{Fujiwara2011}. It thus appears that the afterpulsing effect is not intrinsic to the SNSPD itself. We confirmed experimentally that afterpulsing does indeed disappear when we replace the aforementioned RF Bay amplifiers with a MITEQ AM-1431 amplifier which has a much better low frequency response.

We also want to emphasize the significance of afterpulsing and the recovery profile in QKD applications. One unwanted effect would be if someone chose to operate the system at around $5$~MHz repetition rate (corresponding to about $200$~ns between pulses), and thereby significantly increasing the effective dark count rate, and hence lowering the secure key rate. It is also natural to ask whether such a recovery profile can cause any security concern for a QKD system. First of all, it was shown in \cite{Fung2009} that the proven secure key rate is much lower, in view of detection efficiency mismatch. Second, if Eve always sends signals separated by $180$~ns to Bob, then she can gain side information about the sifted key.

One should be careful about the electronics used in a QKD system as electronics may give false detection signals and afterpulsing effects that may severely undermine security and/or performance of a QKD system. However, even if there is an exploitable loophole due to afterpulsing, the newly developed MDI-QKD \cite{Lo2012} can close all such loopholes in the detection system, including unidentified ones.

\begin{acknowledgments}
Support of the funding agencies CIFAR, CIPI, Connaught, CRC, MITACS, NSERC, QuantumWorks and the University of Toronto is gratefully acknowledged. RHH is supported by a Royal Society University Research Fellowship and the UK EPSRC (grant EP/F048041/1). We thank the Heriot-Watt University workshop for constructing the parts of the SNSPD detector system, and Chandra Mouli Natarajan for his assistance in assembling and testing the SNSPD system. We thank Dr Shigehito Miki at NICT, Japan for providing the SNSPD device chips used in this study. We also thank Li Qian for helpful discussions. Finally, we thank Stephen Julian and his Ph. D. student Fazel Tafti for the use of, and help with their vacuum pump.
\end{acknowledgments}


%

\newpage
\appendix
\setcounter{figure}{0}

\section{Circuit model and simulation}
\label{Section1}

A simple model to describe photodetection in an SNSPD is as follows. A photon striking and being absorbed in the nanowire of the SNSPD will cause a `hotspot' to form at the point of absorption \cite{Goltsman2001r}. The current density around the hotspot will exceed the critical value, and a whole segment of the nanowire will undergo a transition from a superconducting to a resistive state. The resulting voltage pulse can then be amplified and detected. We should point out that this hotspot model is oversimplified. For more details, see Ref \cite{Semenov2005r, Semenov2008r, Bulaevskii2012r}.

To help understand the processes and time scales involved, it is useful to include a circuit model of our SNSPD.

\begin{figure} [!h]
\center
\resizebox{8cm}{!} {
\includegraphics{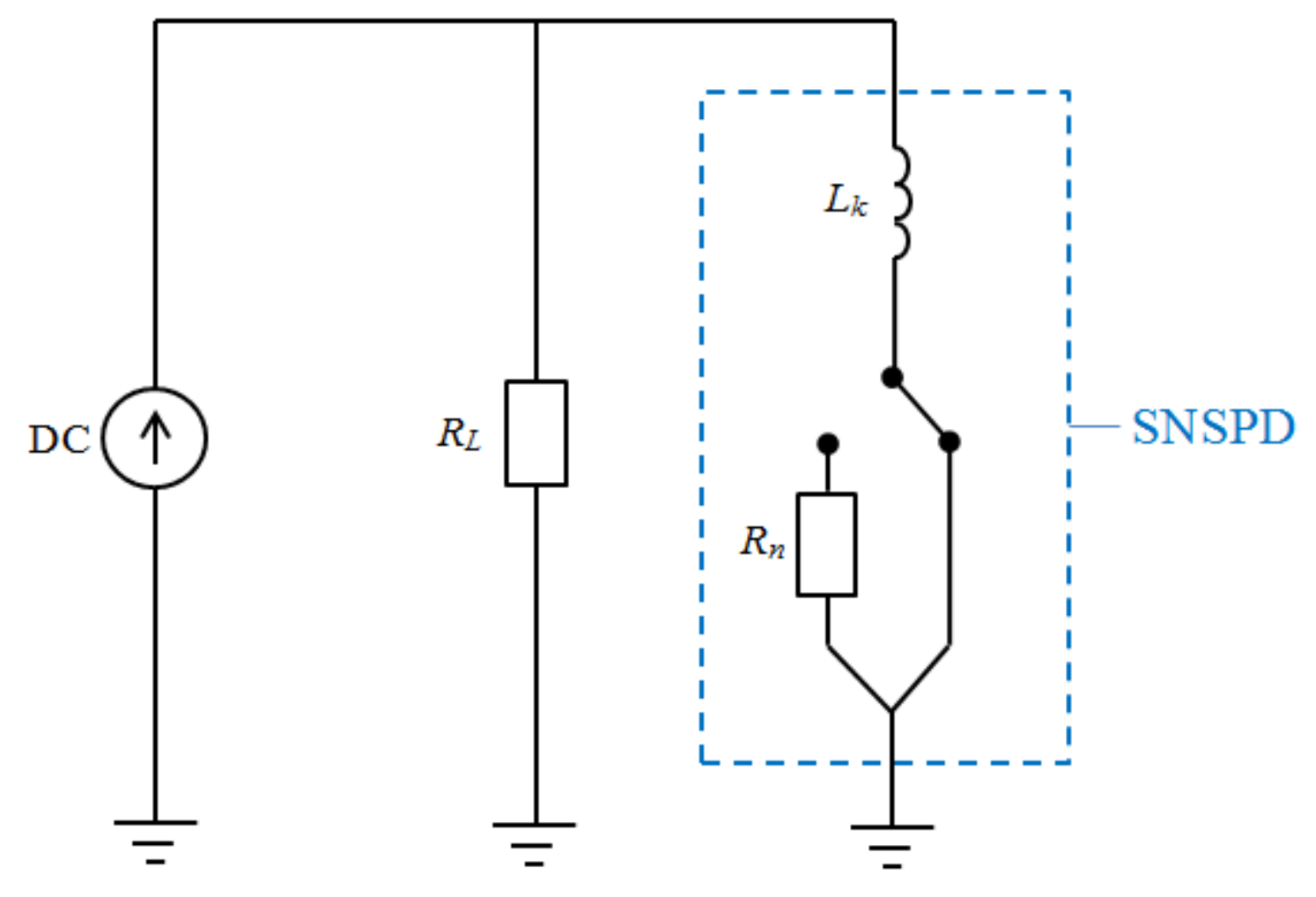} }
\caption{Simple circuit model for an SNSPD \cite{Kerman2006r}. $L_k$ is the kinetic inductance of the superconducting nanowire and $R_n$ is the hotspot resistance of the SNSPD. The absorption of a photon is simulated by opening and closing the switch.}
\label{fig: snspdmodel}
\end{figure}

A simple model for an SNSPD is shown in Fig.~\ref{fig: snspdmodel}.

We have a DC bias source, a load resistor $R_L$, and the SNSPD nanowire. The DC source corresponds to both the voltage source and the $100$~k$\Omega$ resistor in our setup. The load resistor $R_L$ represents the combination of the impedance of the read-out circuitry ($50$~$\Omega$) in parallel with the shunt resistor (also $50$~$\Omega$), giving a combined value of $25$~$\Omega$. It is the voltage across the load resistor that is used to make the measurement of the output signal. The nanowire is modeled as an inductor in series with a variable resistor. We use a simple binary switch model to describe the resistance; it's either zero, when the nanowire is in the superconducting state, or $R_n$ (on the order of several k$\Omega$) when part of the nanowire is in the resistive state. We should note here that this model is a simplification and that after the hotspot is created, the resistance of the SNSPD is in fact time-dependent \cite{Yang2007r}.

The switch is initially set to the superconducting path, but upon a photon striking the nanowire (or a dark count), the switch instantly turns to the resistive path. As the hotspot shrinks the switch resets back to the superconducting path. The current through the nanowire then recovers back to its initial value, although the speed of this recovery is limited by the kinetic inductance of the nanowire $L_k$. This recovery places a dead time on the SNSPD.  The kinetic inductance of our nanowire is relatively large since it's a large area ($20$~$\mu$m~$\times$~20~$\mu$m) meander. As such, the dead time of our SNSPD (time interval after a click in which another click cannot occur) is of the order of $150$~ns.

The circuit model can help us to understand the output voltage pulse shape from a detection, and the time scale over which the current flowing through the SNSPD changes after a detection. Based on the circuit model we expect the current through the SNSPD to drop quickly (given by $L_k/(R_n + R_L) \approx L_k/R_n$ since $R_n \gg R_L$) right after a detection. The current would then slowly increase back to its pre-detection value of $I_B$, with the time constant $\tau = L_k/R_L$. As such, after a time of 3$\tau$ the current would recover to about 95$\%$ of its pre-detection value. This is what determines the detector dead time, and limits the maximum counting rate.

\begin{figure} [!h]
\center
\resizebox{7cm}{!} {
\includegraphics{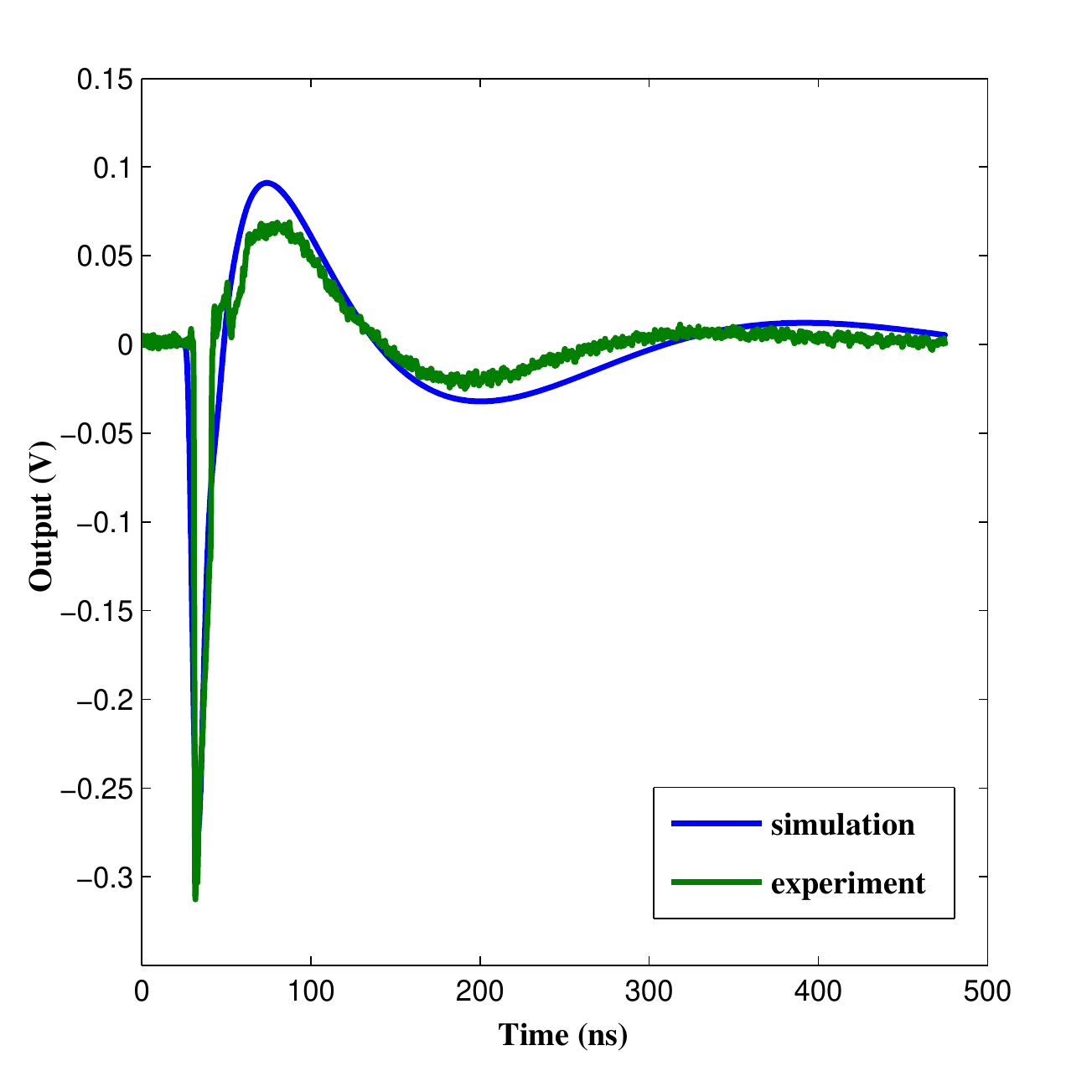} }
\caption{Output pulse shape of a single detection click, starting at $30$~ns, as observed on an oscilloscope (rugged green line), compared to our simulation (smooth blue line). Simulation parameters are $L_k = 500$ nH, $R_n = 5$ k$\Omega$, $R_L = 25$ $\Omega$, $I_b = 25$ $\mu$A. The observed shape shown is the average of ten pulse shapes to reduce appearance of random noise, but the individual pulse shapes are nearly identical to each other.}
\label{fig: pulseSimulation}
\end{figure}

A limitation of the model is that it does not take into account the limited bandwidth of our amplifiers. The shape of the output voltage pulse from a single detection event after it passes through the amplifiers is distorted due to the imperfect frequency response of these amplifiers. While the pulse shape in the absence of bandpass filtering is a pure exponential rise, with filtering, the actual output pulse shape exhibits overshoot and oscillation, and a distortion in the time scale of the pulse. Our simulation in Simulink indicates that it is the poor low frequency response of the amplifiers that causes the overshoot, and we can roughly reproduce the actual output pulse shape by using a $4^{th}$ order bandpass Butterworth filter with passband from $15$~MHz to $580$~MHz (corresponding to the effective frequency range of the first amplifier (LNA-580) in our chain). The effect of restricting the bandwidth in the frequency domain will result in a spread of the pulse shape (and overshoot) in the time domain. See Fig.~\ref{fig: pulseSimulation}. The output voltage pulse coming from the amplifiers of a single detection click, over a time span of $500$~ns, is also shown in Fig.~\ref{fig: pulseSimulation}.

$R_n$ was inferred from the rise time of the pulse and the specific value of 5 kΩ was chosen to best fit our simulation to our experimental data.  There is a fair, but not exact, agreement between the values of $L_k$ and $R_n$ that we use in our simulation and the measured values for this precise type of SNSPD devices in other studies \cite{Miki2008r, OConnor2011r}. It is certainly possible that there is a large uncertainty in these measurements (owing to differences in temperature of measurement, variations in nanowire properties and others).

\section{Latching}
\label{Section2}

Latching occurs when the Joule heating produced in the resistive part of the nanowire exactly balances the cooling, so that the nanowire stays in the resistive state indefinitely. The set-up with a shunt resistor in parallel with the nanowire prevents latching from happening by diverting the current to the shunt and allowing the nanowire to cool off and thus reset \cite{Hadfield2005r}.

In the previous work of Hadfield and coauthors implementing SNSPDs in QKD demonstrations \cite{Hadfield2006r,Takesue2007r}  a shunt resistor was employed to avoid latching in long running QKD experiments.  The alternative is to manually cycle the bias current if the device latches.  Since that time a wider variety of SNSPD designs have become available and studies have been carried out of latching behaviour (notably \cite{Annunziata2010r, Kerman2009r}).  There is an interplay between the embedding impedance, the kinetic inductance of the nanowire and the dynamics of the hotspot.  For the devices used in our study (first reported by \cite{Miki2008r}) the use of a shunt resistor is a pragmatic precaution.  At low current bias it is arguably unnecessary (for long distance QKD where minimizing the dark count rate is the defining factor in determining the limiting quantum bit error rate).  In the high bias regime we have studied here (which is relevant to high bit rate short-haul QKD where the efficiency should be maximized) the shunt resistor is crucial to avoid latching, as observed in our lab.

\section{Estimating afterpulsing contribution in Section VI}
\label{Section3}

In Section VI ``Detection efficiency recover", we measure the detection efficiency recovery of our SNSPD. We do this by sending in a series of double pulses in 2000 ns windows, at fixed intervals ranging from 80 ns to 1000 ns. We calculate the detection efficiency for the second pulse in the pair by taking the ratio of the number of cases where both pulses were detected to the number of cases where only the first pulse was detected.
We do this by plotting a histogram of the time interval between the clicks caused by the first laser pulse and clicks caused by the second laser pulse, combining all the 2000 ns windows making up the experimental run. Sometimes, however, the detections corresponding to the second laser pulse would in fact be due to an afterpulse from the first detection, especially right around the 180 ns mark. We need to subtract the number of these afterpulse clicks from the total count of clicks in the time bin corresponding to the second laser pulse.
To estimate this afterpulse contribution we look at the two neighboring bins either side of the bin where the laser pulses fall, and make a linear extrapolation (i.e. an average of the two neighboring bin values). We then subtract this estimate from the total count in the `laser pulse' bin, giving the estimated number of clicks caused purely by the actual laser pulses. The bins of the histogram are chosen to be small ($<$~5 ns) so that the linear estimate is as accurate as possible, but not so small that the laser pulse contribution does not fall into one bin.


%

\end{document}